\documentclass[prl, preprint, superscriptaddress]{revtex4-1}

\usepackage{amsmath}    
\usepackage{graphicx}   
\usepackage{verbatim}   
\usepackage{color}      
\usepackage[bookmarks=false]{hyperref}

\begin{document}

\title{Effects of contact-line pinning on the adsorption of nonspherical colloids at liquid interfaces}

\author{Anna Wang}
\affiliation{Harvard John A. Paulson School of Engineering and Applied Sciences, Harvard University, Cambridge MA 02138 USA}
\author{W. Benjamin Rogers}
\affiliation{Harvard John A. Paulson School of Engineering and Applied Sciences, Harvard University, Cambridge MA 02138 USA}
\affiliation{Martin Fisher School of Physics, Brandeis University, Waltham MA
  02453 USA}
\author{Vinothan N. Manoharan}
\affiliation{Harvard John A. Paulson School of Engineering and Applied Sciences, Harvard University, Cambridge MA 02138 USA}
\affiliation{Department of Physics, Harvard University, Cambridge MA
  02138 USA}

\email{vnm@seas.harvard.edu}

\begin{abstract}
  The effects of contact-line pinning are well-known in macroscopic
  systems, but are only just beginning to be explored at the microscale
  in colloidal suspensions. We use digital holography to capture the
  fast three-dimensional dynamics of micrometer-sized ellipsoids
  breaching an oil-water interface. We find that the particle angle
  varies approximately linearly with the height, in contrast to results
  from simulations based on minimization of the interfacial energy.
  Using a simple model of the motion of the contact line, we show that
  the observed coupling between translational and rotational degrees of
  freedom is likely due to contact-line pinning. We conclude that the
  dynamics of colloidal particles adsorbing to a liquid interface are
  not determined by minimization of interfacial energy and viscous
  dissipation alone; contact-line pinning dictates both the timescale
  and pathway to equilibrium.
\end{abstract}

\maketitle

The adsorption of a microscopic colloidal particle to a liquid interface
is a dynamic wetting process: in the reference frame of the particle,
the three-phase contact line moves along the particle's surface. Our
understanding of analogous wetting processes in macroscopic systems is
based on two types of models: those that relate the motion of the
contact line to the viscous dissipation in the fluid ``wedge'' bounded
by the solid and the liquid interface~\cite{huh_hydrodynamic_1971,
  dussan_motion_1974, de_gennes_wetting:_1985}, and those that consider
dissipation local to the contact line~\cite{blake_kinetics_1969,
  blake_physics_2006}. Models in this second class treat the motion of
the contact line as a thermally activated process, in which the contact
line transiently pins on nanoscale defects and hops between them. The
two viewpoints are not incompatible: both types of dissipation can be
relevant to experiments~\cite{brochard-wyart_dynamics_1992}, and
thermally activated motion of the contact line can be viewed as a model
for how slip occurs at the solid boundary~\cite{snoeijer_moving_2013}.
The nanoscale defects responsible for pinning the contact line are now
understood to be important in macroscopic wetting experiments at small
capillary number~\cite{prevost_thermally_1999, rolley_dynamics_2007,
  snoeijer_moving_2013} and in phenomena related to contact-angle
hysteresis~\cite{giacomello_wetting_2016}.

\begin{figure}[h]
\centering
  \includegraphics{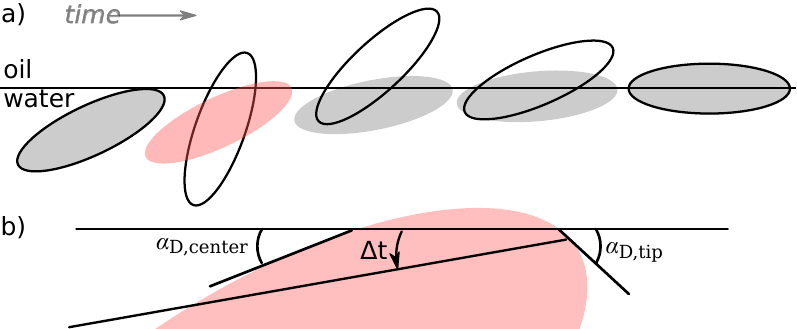}
  \caption{a) Schematic showing the cross-section through the long axis
    of an ellipsoidal particle breaching an interface. Simulations
    assuming interfacial energy is minimized at each time step predict
    trajectories (solid lines) that are non-monotonic in both
    center-of-mass position and polar
    angle~\cite{gunther_timescales_2014, degraaf_adsorption_2010}. Our
    experimental results (shaded ellipses) show a monotonic increase in
    both height and polar angle. See also the movie in the Supplemental
    Material~\cite{Note1}. b) Schematic of contact-line motion for one
    scenario in a). A line shows where the contact line moves in the
    frame of reference of the particle after time $\Delta$t. The dynamic
    contact angles at two points are shown.}
  \label{fig:cartoons}
\end{figure}

Such defects---and the associated hopping of the contact line---have
also proven important for understanding the dynamics of microscopic
colloidal particles at liquid interfaces. These particles can strongly
adhere to the interface, owing to the large change in the total
interfacial energy of the system once the particles
bind~\cite{binks_colloidal_2007}. The resulting systems are used to
study phase transitions and self-assembly in two
dimensions~\cite{pieranski_two-dimensional_1980,
  aveyard_compression_2000, zahn_dynamic_2000, bausch_grain_2003,
  koenig_experimental_2005, cavallaro_curvature_2011,
  botto_capillary_2012}, to fabricate new
materials~\cite{paunov_supra_2004, retsch_nanofab_2009,
  isa_particle_2010}, and to formulate Pickering emulsions and
colloidosomes~\cite{ramsden_separation_1903,
  pickering_spencer_umfreville_emulsions_1907,
  dinsmore_a.d._colloidosomes:_2002, mcgorty_colloidal_2010}. But nearly
all particles used in such systems have nanoscale surface features that
can pin the contact line~\cite{wang_contact_2016}. Whereas strong
pinning sites can affect interactions~\cite{kralchevsky_particles_2001}
and motion on curved interfaces~\cite{sharifi-mood_curvature_2015}, even
weak pinning sites can dramatically affect dynamics. For example,
Boniello and coworkers~\cite{boniello_brownian_2015} showed that the
in-plane diffusion of particles straddling an air-water interface is
inconsistent with models of viscous drag in the bulk fluids but
consistent with models of contact-line fluctuations caused by nanoscale
defects. Also, Kaz, McGorty, and coworkers~\cite{kaz_physical_2012}
showed that spherical colloidal particles breaching liquid interfaces
relax towards equilibrium at a rate orders of magnitude slower than that
predicted by models of viscous dissipation in a fluid wedge. Dynamic
wetting models based on thermally activated
hopping~\cite{blake_kinetics_1969, colosqui_colloidal_2013} fit the
observed adsorption trajectories well over a wide range of timescales.

Here we examine the effects of transient pinning and depinning on the
\emph{pathway} that ellipsoidal colloidal particles take to equilibrium,
and not just the time required to get there. By ``pathway'' we mean the
way in which the degrees of freedom vary with time. For example, a
spherical particle has one degree of freedom---its height relative to
the interface. Its pathway to equilibrium does not depend on pinning
effects: at low Reynolds number, the height changes monotonically with
time, regardless of whether the contact line becomes pinned along the
way. By contrast, an ellipsoid of revolution, or spheroid, has two
degrees of freedom---its center-of-mass position and orientation---which
need not vary monotonically with time. In fact, molecular
dynamics~\cite{gunther_timescales_2014} and Langevin dynamics
studies~\cite{degraaf_adsorption_2010} predict that both the height and
orientation of ellipsoidal particles breaching an interface vary
non-monotonically with time (Figure~\ref{fig:cartoons}a). These
simulations assume that the particles follow a quasi-static approach to
equilibrium, such that the total interfacial energy is minimized at each
step along the way, and they do not model contact-line pinning. They
predict an equilibration time of 10 $\mu$s based on the viscous
dissipation, but previous experimental studies of prolate spheroids at
interfaces by Coertjens \textit{et al.}~\cite{coertjens_contact_2014}
and Mihiretie \textit{et al.}~\cite{mihiretie_mechanical_2013} suggest
that the equilibration time is much longer, hinting at pinning effects.

We use a fast 3D imaging technique, digital holographic microscopy, to
observe ellipsoidal (prolate spheroidal) colloidal particles adsorbing
to an oil-water interface. We find that not only is the equilibration time orders of magnitude
larger than that predicted by simulations, but the position and
orientation of the particle vary \emph{monotonically} with time, in
contrast to the nonmonotonic pathways found in simulations
\cite{degraaf_adsorption_2010, gunther_timescales_2014}. Interestingly,
the center-of-mass position and the polar angle of the particle are
coupled, so that the particle ``rolls'' into its equilibrium position as
if it were subject to a tangential frictional force
(Figure~\ref{fig:cartoons}a and Supplemental Material~\cite{Note1}). We
argue that these effects are due to contact-line pinning.

\begin{figure}[h]
\centering
  \includegraphics{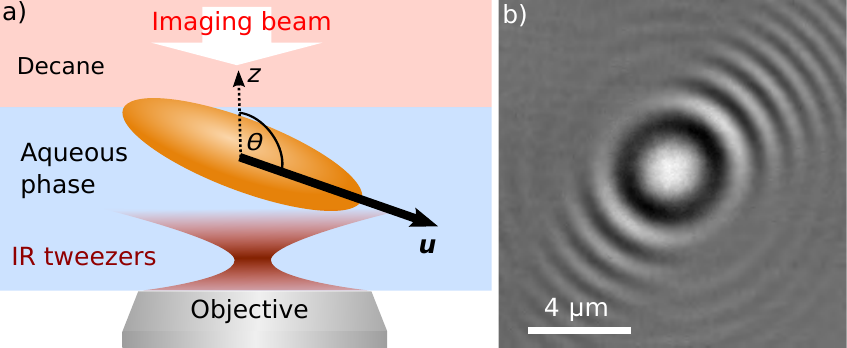}
  \caption{Schematic of experimental setup. a) The sample is illuminated
    with a red collimated laser ($\lambda$ = 660 nm). A
    counterpropagating infrared laser ($\lambda$ = 785 nm) is used to
    gently push the particles against gravity towards the interface. In
    the coordinate system used here, the imaging axis lies along $z$,
    and the interface is at $z$=0. The center-of-mass position and the
    polar angle $\theta$ of the particle are defined relative to the
    laboratory frame by a unit vector $\mathbf{u}$ that points along the
    long axis of the ellipsoid. b) A typical hologram of an ellipsoid at
    an interface. See also the movie in the Supplemental
    Material~\cite{Note1}.}
  \label{fig:dhmsem}
\end{figure}

To make ellipsoidal particles, we heat 1.0-$\mu$m-diameter
sulfate-functionalized polystyrene particles (Invitrogen) above their
glass transition temperature and stretch them~\cite{Note1}. Using the
apparatus shown in Figure~\ref{fig:dhmsem}, we capture holograms of
individual ellipsoids at 100 frames per second as they approach an
interface between decane and a water/glycerol mixture. The holograms
encode the three-dimensional position and orientation of the particle in
the spacing and shape of the interference fringes. We extract this
information, along with the particle size and refractive index, by
fitting a T-matrix model~\cite{mishchenko_capabilities_1998} of the
scattering from the particles to our data~\cite{Note1}.

\begin{figure}[h]
\centering
  \includegraphics{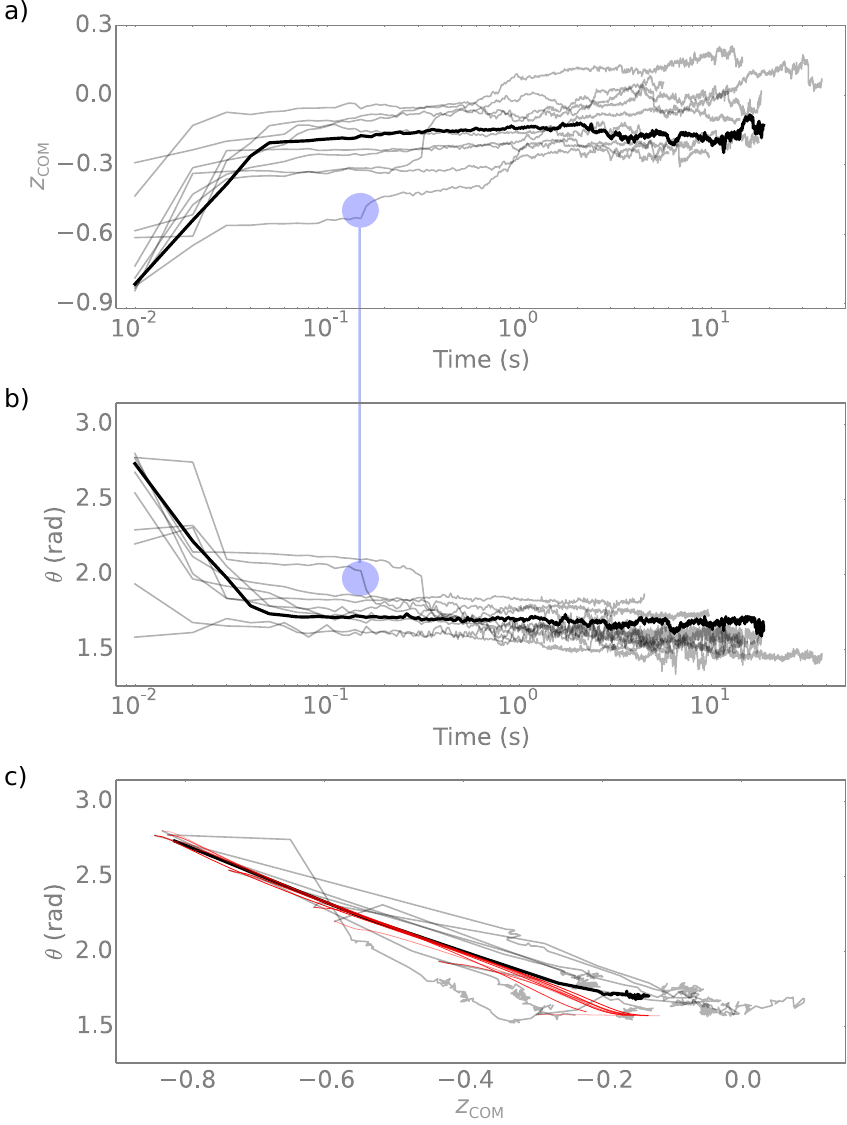}
  \caption{The height (a) and polar angle $\theta$ (b) of ellipsoids
    after they breach (breach at $t$ = 0 s) appear correlated (example
    marked with blue circles). To compare data from particles of
    different aspect ratios on the same plot, we normalize the
    center-of-mass height relative to the
    interface~\cite{degraaf_adsorption_2010} to obtain $z_\mathrm{COM} =
    z/\sqrt{2a^2+b^2}$ , where $a$ and $b$ are the semi-minor and
    semi-major axes of the spheroid. c) The polar angle varies with the
    height. Our model (red), based on contact-line hopping over defects,
    produces almost linear relationships between $z$ and $\theta$ for
    the particles from the experiments.}
  \label{fig:thetaz}
\end{figure}

We find that the particles relax to equilibrium slowly, despite some
abrupt motion along the way, as shown in Figure~\ref{fig:thetaz}a. After
1 s, or five orders of magnitude longer than the equilibrium time
observed in simulations~\cite{degraaf_adsorption_2010}, the particles
are still not equilibrated. Furthermore, their height scales roughly
linearly with the logarithm of the elapsed time after the breach.
Because this is the same scaling observed by Kaz and
coworkers~\cite{kaz_physical_2012}, the likely origin of the slow
dynamics is contact-line pinning and depinning, the same mechanism
observed in spherical particles. The slow dynamics are perhaps
unsurprising, since the particles are stretched versions of those used
by Kaz \emph{et al.}~\cite{kaz_physical_2012}.

More surprisingly, we find that the abrupt changes in height
(Figure~\ref{fig:thetaz}a) correlate with abrupt changes in the polar
angle $\theta$ (Figure~\ref{fig:thetaz}b). Indeed, when we examine
$\theta$ as a function of $z$ we find that the relationship is
approximately linear (Figure~\ref{fig:thetaz}c). Furthermore, although
the particles approach the interface from a variety of different polar
angles, the $\theta$--$z$ plots form lines with similar slopes, hinting
at the presence of a dynamical attractor.

The linear relationship between $\theta$ and $z$ is reminiscent of
rolling, where translation and rotation are coupled by friction. The
particles ``pivot'' into the interface, as shown by the rendering of
individual points along the observed trajectories
(Figure~\ref{fig:cartoons}a and Supplemental Material~\cite{Note1}). In
contrast, the simulations by G\"unther \textit{et
  al.}~\cite{gunther_timescales_2014} and de Graaf \textit{et
  al.}~\cite{degraaf_adsorption_2010} predict that $\theta$ and $z$
should vary non-monotonically in time (Figure~\ref{fig:cartoons}a).

We therefore seek a different model to explain the observed
rotational-translational coupling, one that takes into account
contact-line pinning. We adopt the viewpoint of Kaz \textit{et
  al.}~\cite{kaz_physical_2012} and assume that the contact line pins to
defects on the surface of the particle and hops between pinning sites
with the aid of thermal kicks. We calculate the velocity of the contact
line using an Arrhenius equation coupled to a model for the force,
determined by the dynamic contact angle $\alpha_\mathrm{D}(t)$. In
contrast to the model for spheres, our model allows
$\alpha_\mathrm{D}(t)$ to vary as a function of position on the
particle, though we still assume that the interface remains flat at all
times, a simplification that we justify based on the energetic cost of
bending the interface.

For a horizontal interface defined by a denser aqueous phase on the
bottom and an oil phase on top (Figure~\ref{fig:dhmsem}), the force on
each segment of the contact line is then determined from the imbalance
between the three interfacial tensions (oil--water $\sigma_\mathrm{ow}$,
particle--water $\sigma_\mathrm{pw}$, and particle--oil
$\sigma_\mathrm{po}$):
\begin{equation}
\begin{split}
  F_\mathrm{cl} & = {\sigma}_\mathrm{ow}\cos{\alpha}_\mathrm{D}(t)+\sigma_\mathrm{pw}-\sigma_\mathrm{po} \\
  & = {\sigma}_\mathrm{ow}\left(\cos{\alpha}_\mathrm{D}(t)-\cos{\alpha}_\mathrm{E}\right)
\end{split}
\label{f}
\end{equation}
where $\alpha_\mathrm{E}$ is the equilibrium contact angle. To
understand this relation, consider Figure~\ref{fig:cartoons}b. The force
is tangent to the particle. Because the contact angle is defined
relative to the aqueous phase, $F_\mathrm{cl}$ is positive and the
contact line moves \emph{down} the particle (in the particle frame) if
the particle approaches the interface from the aqueous phase
(${\alpha}_\mathrm{D} < {\alpha}_\mathrm{E}$). If the particle
approaches the interface from the oil phase (${\alpha}_\mathrm{D} >
{\alpha}_\mathrm{E}$), $F_\mathrm{cl}$ is negative and the contact line
moves \emph{up} the particle. Because the particle is nonspherical, the
force per unit length along the contact line is asymmetric about the
short axis of the ellipsoid unless the particle breaches the interface
at exactly $\theta$ = $\pi$/2.

The direction and magnitude of this force determine the velocity of the
contact line along the particle surface. We relate $F_\mathrm{cl}$ to
the velocity of the contact line using an Arrhenius equation that is
valid when forward hops dominate~\cite{blake_kinetics_1969,
  kaz_physical_2012}
($\lvert\alpha_\mathrm{E}-\alpha_\mathrm{D}\rvert\gtrsim 0.01$, as
discussed in the Supplemental Material~\cite{Note1}):

\begin{equation}
V = \frac{F_\mathrm{cl}(t)}{\lvert F_\mathrm{cl}(t) \rvert} V_0 \exp \left( -\frac{U}{kT}+\frac{\lvert F_\mathrm{cl}(t)\rvert A}{2kT} \right)
\label{velocity}
\end{equation}
where $V$ is the velocity of the contact line tangent to the particle,
$V_0$ is a molecular velocity scale, $A$ is the area per defect on the
surface of the particle, and $U$ is the energy with which each defect
pins the contact line. When spherical particles start in the aqueous
phase, Equation~\ref{velocity} reduces to the form described in the
Supplementary Information of Kaz \textit{et
  al.}~\cite{kaz_physical_2012}. Further details of the model are given
in the Supplemental Material~\cite{Note1}.

Before comparing the model to the data, we first determine which
parameters in the model control the trajectory. The dynamic contact
angle is greater near the tips of the particle than it is near the
center ($\alpha_\mathrm{D,tip} > \alpha_\mathrm{D,center}$), as
illustrated in Figure~\ref{fig:cartoons}b. The part of the contact line
nearest the tip travels more slowly than the part nearest the center
($V_\mathrm{tip} < V_\mathrm{center}$ according to Equations~\ref{f}
and~\ref{velocity}), leading to the observed pivoting motion.
Considering only these two points, the ratio
$V_\mathrm{center}$/$V_\mathrm{tip}$ =
$\exp\left(\sigma_\mathrm{ow}A\left(\cos \alpha_\mathrm{D,center} - \cos
    \alpha_\mathrm{D,tip}\right)/2kT\right)$ approximates the rate at
which $z$ and $\theta$ change relative to each other. We therefore
expect the shape and aspect ratio of the particle, which determine the
values of $\alpha_\mathrm{D}$ for a given $\theta$ and $z$, and the area
per defect $A$ to control the form of the $\theta$--$z$ curve. Changing
the size of the particle, $U$, or $\alpha_\mathrm{E}$ alters how
$\theta$ and $z$ depend on time, but not how $\theta$ and $z$ evolve
with each other. We explore the effect of particle shape further below.

The model produces trajectories that agree with experimental
observations. In Figure~\ref{fig:thetaz}c, we plot calculated
$\theta$-$z$ trajectories for each of the particles using the measured
aspect ratio determined from fitting the holograms and a defect area
equal to that measured in Kaz \textit{et al.}~\cite{kaz_physical_2012},
$A$ = 4 nm$^2$. We have strong evidence that $A$ is nanoscale, as
discussed in the Supplemental Material~\cite{Note1}. The modeled
$\theta$ and $z$ are both monotonic with time, in contrast to the
predictions from earlier simulations. Moreover, $\theta$ varies linearly
with $z$, reproducing the translational-rotational coupling in our
experimental results. The slope predicted by the model (-1.75 $\pm$ 0.16
rad) agrees with the average slope in our experimental results (-1.64
$\pm$ 0.76 rad).

Our model predicts that all trajectories collapse onto one line for a
given $A$ (Figure~\ref{fig:thetaz}c). de Graaf and coworkers found a
similar ``dynamical attractor'' for ellipsoids in their Langevin
simulations~\cite{degraaf_adsorption_2010}, while G\"unther \textit{et
  al.}~\cite{gunther_timescales_2014} found that the adsorption
trajectory depends sensitively on the angle of the particle when it
first touches the interface. Our model predicts an attractor, but the
attractor arises from the pinning, and in particular from how pinning
ensures that the contact-line velocity near the tip is always smaller
than that near the center.

The only discrepancy between the model and our data is that the slopes
of the modeled trajectories are more narrowly distributed than the
experimentally observed slopes. This discrepancy may be due to
deviations in shape from perfect prolate spheroids (see image of the
particles in the Supplementary Material~\cite{Note1}), which would
affect the curvature of the particle and hence $\alpha_\mathrm{D}$. The
area per defect $A$ might also vary between particles. Our data falls
between the calculated attractors for particles with $A$ = 1 nm$^2$ and
30 nm$^2$. Finally, the stretching of the particles might lead to an
inhomogeneous defect density, which we do not account for in our model.
However, the agreement between the average observed and calculated
slopes suggest that our assumption of a uniform defect density is valid
to within the uncertainties of our measurements~\cite{Note1}.

\begin{figure}[h!]
\centering
  \includegraphics{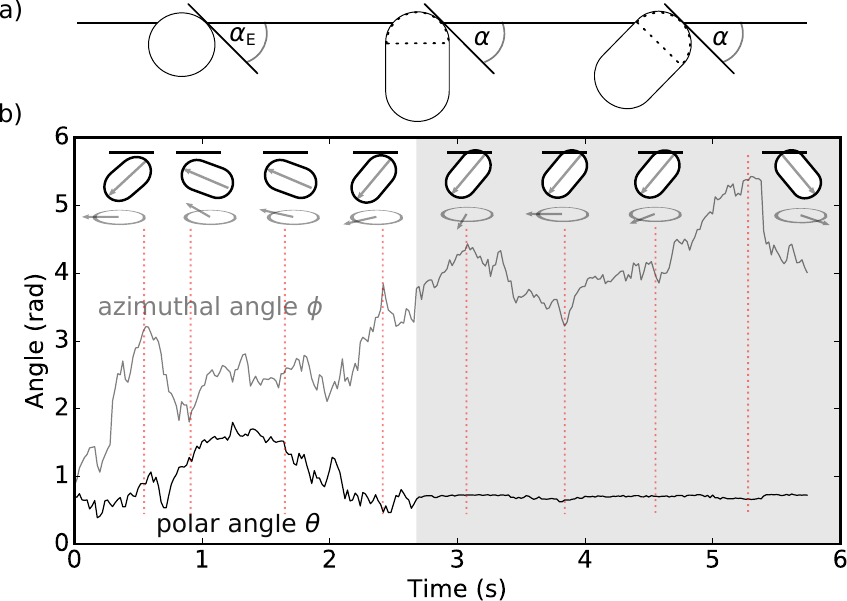}
  \caption{a) A hydrophilic spherocylinder ($\alpha_\mathrm{E} <$
    $\pi$/4; hemispherical caps drawn in dotted lines) can reach a local
    minimum in the interfacial energy when the contact angle $\alpha$ =
    $\alpha_\mathrm{E}$. b) Experimental data showing that a hydrophilic
    silica spherocylinder freely rotates about the $z$-axis once it has
    attached to the interface (shaded region), but its polar angle
    remains fixed. Arrows indicate the particle's polar and azimuthal
    angles at the time points indicated by the red dotted lines.}
  \label{fig:spherocylinder}
\end{figure}

Our model further predicts that if the particles were spherical, there
would be no net torque on the particle, owing to the symmetry. To verify
this prediction, we examine spherocylindrical particles
(Figure~\ref{fig:spherocylinder}), each of which consists of a cylinder
with hemispherical caps. If the spherocylinder is hydrophilic, such that
a sphere of the same material has an equilibrium contact angle smaller
than $\pi/4$ (Figure~\ref{fig:spherocylinder}a), it can attach to the
interface at a range of polar angles all having the same energy
(Figure~\ref{fig:spherocylinder}a). Our model predicts that the contact
line will not exert a torque on the particle and that the particle
should therefore remain at the same polar angle.

Indeed, we observe that when hydrophilic spherocylinders breach the
interface, their polar angle remains fixed, although their azimuthal
angle can still fluctuate (Figure~\ref{fig:spherocylinder}b). The
absence of equilibration is surprising because the minimum-energy
configuration for spherocylinders is $\theta$ = $\pi/2$. However, the
observed trajectories agree with the prediction of our model.

We therefore conclude that accounting for contact-line pinning, and not
just interfacial energy, is necessary to understand the dynamics of
nonspherical particles at liquid interfaces. The slow relaxation
observed in such systems is just one manifestation of the pinning; as we
have shown here, the pinning can also alter the pathway to equilibrium.
The agreement between model and experiment suggests that the pathway is
controlled not only by the size of the defects on the solid surfaces,
but also by the local curvature of the particle. Thus, not only is the
road to equilibrium long, but it also depends on the details of the
shape of the particle.

These results have practical implications for assembling particles at an
interface and, at the same time, lead to new fundamental understanding
of how dynamic wetting influences particles at interfaces. In terms of
practical implications, the unexpectedly long adsorption times we find
in our experiments might significantly affect the aging of Pickering
emulsions. Although we have neglected the curvature of the interface in
our simple dynamical model, ellipsoidal particles should induce a
quadrupolar capillary field in equilibrium~\cite{loudet_capillary_2005}.
Therefore the equilibration of multiple particles at the interface might
be complicated by the slowly evolving capillary interactions between
them. Also, the shape of the particle might play a large role in
determining if and how such systems arrive at equilibrium. Particles
like the spherocylinder (and related particles such as \textit{E. coli}
and other pill-shaped bacteria), can get stuck at a particular angle
after attaching to the interface because the contact line ``sees'' a
sphere (Figure~\ref{fig:spherocylinder}a). Such particles would
therefore need a large thermal fluctuation to rotate toward their
equilibrium configuration.

In terms of fundamental understanding, the ability of our model to
recreate the nearly linear $\theta$--$z$ adsorption trajectories for
ellipsoids validates the idea that contact-line pinning couples
orientational and translational degrees of freedom. The role that
pinning plays is akin to the role of friction in rolling. Here, however,
the friction arises not from the interactions between microscopic
features on two solid surfaces, but from the interactions between
nanoscale defects on a solid surface and a deformable liquid--liquid
interface. Though these interactions are weak enough to be disrupted by
thermal fluctuations, and the capillary driving forces are large, the
frictional coupling can nonetheless drive adsorption trajectories that
are observably different from those expected from interfacial energy
minimization and viscous dissipation.

\begin{acknowledgments}
  We thank Thomas E. Kodger and Peter J. Yunker for their apparatus and
  advice for making ellipsoids, Michael I. Mishchenko for allowing us to
  incorporate his T-matrix code with our hologram analysis software
  HoloPy, Christopher Chan Miller for help with adapting Fortran code
  for use with HoloPy, and Kundan Chaudhary for synthesizing the silica
  spherocylinders. This work was funded by the National Science
  Foundation (NSF) through grant no. DMR-1306410 and by the Harvard
  MRSEC through NSF grant no. DMR-1420570.
\end{acknowledgments}



\vspace{3cm}


\centerline{\textbf{Effects of contact-line pinning on the adsorption of nonspherical
  colloids}} \centerline{\textbf{  at liquid interfaces: Supplementary Information}}

\subsection*{Making the particles}
\begin{figure}[h]
\centering
  \includegraphics[width=0.45\textwidth]{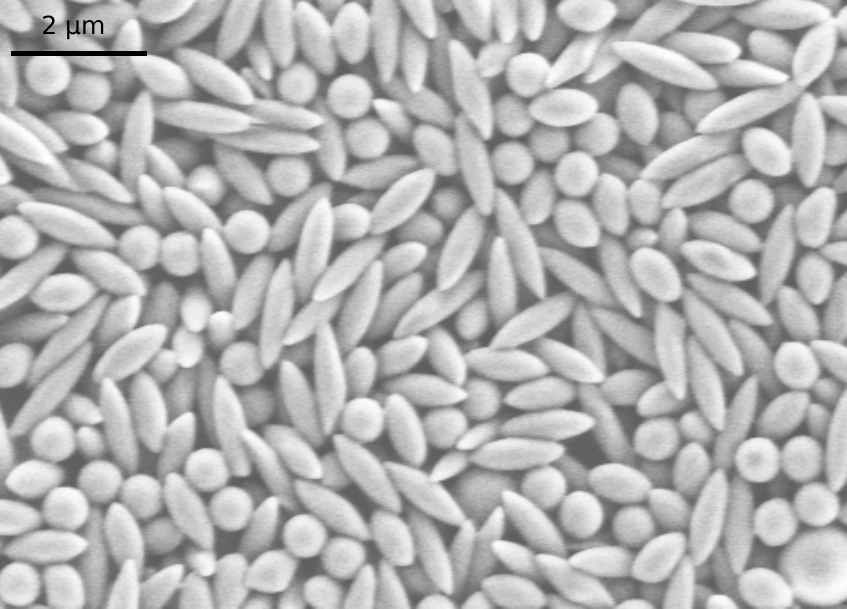}
  \caption{Scanning electron microscopy image of the stretched spheres. Most
    particles are approximately prolate spheroids.}
  \label{fig:sem}
\end{figure}

We prepare ellipsoidal particles by following a previously published
protocol~\cite{ho_preparation_1993} with slight modifications. We embed
spherical particles in a 10\% w/w poly(vinylalcohol) matrix (average M$_w$
146,000-186,000, 87-89\% hydrolyzed from Sigma-Aldrich). We then allow the film
to dry until brittle over at least one week at room temperature. Afterward, we
clamp the film in a custom-made apparatus consisting of two clamps on rails and
heat the polymer film with a heat gun (Master Appliance, set to 1000 $^\circ$F)
held approximately 5 cm over the film. After heating the film for 30 s, we
stretch the film by pulling apart the two clamps~\cite{ho_preparation_1993,
  yunker_suppression_2011}. We dissolve the film in a mixture of isopropyl
alcohol ($\geq$99\% purity, BDH chemicals) and deionized water (Elix, EMD
Millipore, resistivity 18.2 M$\Omega \cdot$cm), then wash the freed particles in
fresh isopropanol/water and then pure deionized water using a ``double
cleaning'' protocol as described in Coertjens \textit{et
  al.}~\cite{coertjens_contact_2014}. The starting particles are
sulfate-functionalized polystyrene spheres ($d$ = 1.0 $\mu$m) from Invitrogen.
The stretched particles are shown in Figure~\ref{fig:sem}.

The silica spherocylinders are synthesized using a modified one-pot
method~\cite{chaudhary_janus_2012, kuijk_synthesis_2011}, and their size (as
determined by scanning electron microscopy) is $1 \pm 0.2$ $\mu$m by $2 \pm
0.2$ $\mu$m.

We dilute the particles to a volume fraction of less than 10$^{-5}$ in an
aqueous solution of glycerol (59\% w/w) in water so that in the experiments,
there is no more than one particle in the 120 $\times$ 120 $\times$ 100 $\mu$m
imaging volume. We place approximately 5 $\mu$L of suspension in a custom
sample holder and add approximately 300 $\mu$L of anhydrous decane ($\geq$99\%
purity, Sigma) to form the interface. The aqueous solution is refractive-index
matched to the decane ($n$ = 1.41) to minimize reflections from the interface.

\subsection*{Taking and analyzing holograms}

We capture the dynamics of the ellipsoidal particles with a digital holographic
microscope. Our experimental apparatus is described in detail by Kaz \textit{et
  al.}~\cite{kaz_physical_2012}. In brief, we capture holograms with an in-line
digital holographic microscope built on a Nikon TE2000-E inverted microscope. We
push particles towards the interface with an out of focus optical trap
($\lambda$ = 785 nm, Thorlabs L785P090), while a laser diode ($\lambda$ = 660
nm, Thorlabs HL6545MG) illuminates them. The scattered and unscattered light
interfere to form a hologram, which is collected with a 100$\times$, NA = 1.4
oil-immersion objective (Nikon) with a $n$ = 1.414-index oil to minimize
spherical aberration in the hologram~\cite{kaz_physical_2012,
  ovryn_imaging_2000}. The holograms are then recorded with a Photon Focus
MVD-1024E-160 camera at 50 $\mu$s exposure time.

We analyze the recorded holograms by fitting a light-scattering model to the
data using the HoloPy package (\url{http://manoharan.seas.harvard.edu/holopy}).
We have previously shown that we can use holography and an implementation of the
discrete dipole approximation
(A-DDA~\cite{yurkin_discrete-dipole-approximation_2011}) to track non-spherical
particles diffusing in water with high temporal resolution and spatial
precision~\cite{wang_using_2014}. We use the same technique to analyze holograms
from the spherocylinder. For the ellipsoids, we calculate the scattered field
with a T-matrix solution~\cite{mishchenko_capabilities_1998}, which is faster
than the discrete dipole approximation: calculating a 200 by 200 pixel hologram
of a 0.5- by 2.0-$\mu$m ellipsoid takes 1 s using the T-matrix solution
compared to 10 s with A-DDA. We use the fit results from one frame, obtained
using the Levenberg-Marquardt algorithm, as the initial guess for the following
frame in the time series.

To fit a time-series of holograms, we must determine good initial guesses for
the parameters in the first frame, including the refractive index of the
particle, its size, orientation, and position. For the refractive index we use
that of bulk polystyrene ($n$ = 1.59 at 660 nm). To estimate the orientation and
size of the particles, we calculate reconstructions of the hologram by
Rayleigh-Kirchoff back-propagation~\cite{kreis_frequency_2002}. We estimate the
height from the $z$ position where the reconstruction resembles an in-focus
bright-field image of the particles. We then use this reconstruction slice to
estimate the particle's size and orientation. We estimate the particle's
$x$--$y$ position using a Hough transform based algorithm in HoloPy. These
estimates are then refined in the fit.

\subsection*{Modeling contact-line motion on the surface of a particle}
To model the motion of the contact line along the surface of a spheroid,
we work in the particle's frame of reference and start from an initial
position in which the particle just touches the interface at a given
polar angle $\theta_\mathrm{init}$. To generate a trajectory that can be
compared to experimental data, we choose $\theta_\mathrm{init}$ to be
the observed polar angle of the particle just before it breaches.

The velocity of the contact line along the surface of the particle is given by 
\begin{equation}
V = \frac{F_\mathrm{cl}}{\lvert F_\mathrm{cl} \rvert }V_0 \exp(-U/kT + \lvert F_\mathrm{cl} \rvert A/2kT)
\label{velocity2}
\end{equation}
We define the contact line as the intersection between a plane (the interface)
and the surface of a rotated prolate spheroid. For an interface at $z$=0, the
contact line is given by
\begin{equation*}
y = \pm \sqrt{a^2-\frac{a^2}{b^2} (-x \sin \theta+k \cos \theta)^2-(x \cos \theta + k\sin \theta)^2}
\end{equation*}
where $a$ and $b$ are the minor and major semi-axes of the prolate spheroid,
$\theta$ is the polar angle, and $k$ is the height of the particle. In our
calculations, we use 5000 points to define the contact line.

For each of the points along the contact line, we calculate the velocity
from Equation~\ref{velocity2} using $T$ = 295 K, $V_0$ = 1 $\mu$m/s,
$U$ = 20 $kT$, $\sigma_\mathrm{ow}$ = 37 mN/m, and $A$ = 4 nm$^2$. We
take the value for $U$ from Colosqui \textit{et
  al.}~\cite{colosqui_colloidal_2013}, and the values for
$\sigma_\mathrm{ow}$ and $A$ from Kaz \textit{et
  al.}~\cite{kaz_physical_2012}. We then multiply each velocity by a
time-step $\delta t$ to find the displacements $\delta l$. The $\delta
l$ at each of the 5000 points are in the direction tangent to the
particle. The time-steps we use are smaller in the beginning of the
trajectory than at the end ($\delta t_i$ = $e^{-2}$, $e^{-1.98}$,
$e^{-1.96}$, $...$, $e^{0}$, $...$, $e^{1.98}$, $e^{2}$) because the
driving force $F_\mathrm{cl}$ is larger in the beginning of the
trajectory. Using uniformly small time-steps for the whole integration
does not change the $\theta$--$z$ trajectories, only the speed of the
integration. After each time-step, we update each point along the
contact line with the respective value of $\delta l$, and we fit the
updated points to a plane to define the new position of the interface.
The height $z$ and polar angle $\theta$ are then determined by the angle
and position of the plane relative to the center of the ellipsoid.

\subsection*{Choice of $A$ and how it affects the velocity of the contact line}

When a prolate spheroid breaches the interface, our experiments show
that the contact line moves along the surface of the particle at
different speeds, rotating the particle until it ``lies down'' at the
interface. We calculate the ratio of the velocities of the slowest- to
fastest-moving parts of the contact line using the following equation:
\begin{equation}
\frac{V_\mathrm{center}}{V_\mathrm{tip}} = \exp\left(\frac{\sigma_\mathrm{ow}\left(\cos \alpha_\mathrm{D,center} -\cos \alpha_\mathrm{D,tip}\right)A}{2kT}\right)
\label{eq:ratio}
\end{equation}

The ratio in equation \ref{eq:ratio} is sensitive to the dynamic contact
angles $\alpha_\mathrm{D,center}$ and $\alpha_\mathrm{D,tip}$, which are
set by the shape of the particles, and the value of $A$. Our group
determined $A$ to be approximately 4 nm$^2$ for sulfate polystyrene
particles \cite{kaz_physical_2012}, the value we use in this work, and
1-10 nm$^2$ for a range of other particles \cite{wang_contact_2016}.
Boniello \textit{et al.} \cite{boniello_brownian_2015} found $A$ to be
on the order of 1 nm$^2$ for silica and polystyrene particles.
Contact-line pinning onto nanometer-scale defects has been observed in
macroscopic wetting experiments, and recent DFT simulations have shown
that pinning/unpinning of the contact line on 1-nm-scale surface defects
can result in macroscopic contact angle hysteresis
\cite{giacomello_wetting_2016}. We therefore believe choosing $A$ to be
on the nanometer-scale in our simulations is justified.

If $A$ is much smaller than 1 nm$^2$, then thermal fluctuations will be
comparable to the work needed to move a segment of the contact line from
one defect to the next. In this situation, the contact line will have
more defects to pin it. For such conditions, equation \ref{velocity2}
may no longer provide a valid description of the motion of the contact
line.

\subsection*{The effect of stretching the particles on their surface properties}
The average slopes of the $\theta$-$z$ trajectories for our model and
experimental data agree, despite our assumption that the particles have
uniform surface properties. The quantitative agreement suggests that any
effects from stretching the particles are not detectable to within the
uncertainty of our data. This may be because the surface area of the
particles did not increase very much when they were stretched to an
aspect ratio of between 2.5 and 4: particles stretched to aspect ratio
of 2.5 have 13\% more surface area than their spherical form, and
particles stretched to an aspect ratio of 4 have 28\% more. An
approximately 20-30\% increase in surface area in parts of the particle
is unlikely to alter the adsorption trajectories by much, given that the
attractors for particles with $A$ = 1 and 30 nm$^2$ are not that
different in slope. All of the experimental data falls between these two
attractors.

We also do not expect the stretching to alter the qualitative form of
the trajectories. Because the surface area of the particle increases
more near its center than at its tips, the density of defects near the
center should decrease relative to that near the edges, such that
$A_\mathrm{tip} < A_\mathrm{center}$. Local alignment of polymer chains
during stretching may also occur, leading to enhanced hydrophobicity
near the center relative to the tips~\cite{coertjens_contact_2014}, such
that $\alpha_\mathrm{E,tip} < \alpha_\mathrm{E,center}$. Equation
\ref{velocity2} shows that geometry and surface properties all favor
$V_\mathrm{tip} < V_\mathrm{center}$, resulting in the particle pivoting
as it breaches.

\subsection*{Direction of contact-line hops}
In equation \ref{velocity2} we do not take into account ``backward
hops,'' motion of the contact line in the reverse direction due to
thermal fluctuations. Neglecting backward hops is a valid simplification
in the regime where $\cos \alpha_\mathrm{D}$ is far from $\cos
\alpha_\mathrm{E}$. When $\cos \alpha_\mathrm{D}$ approaches $\cos
\alpha_\mathrm{E}$, thermal fluctuations can result in backward hops,
and the model must be modified. The Supplementary Material of Kaz
\textit{et al.}~\cite{kaz_physical_2012} demonstrates that for the
spherical (unstretched) versions of the particles that we use in this
study, backward hops can be neglected so long as the dynamic contact
angle is more than 0.01 radians from the equilibrium contact angle.
Therefore in this work, we consider only the regime where
forward-hopping dominates.

\subsection*{MOVIE-ellipsoid\_holo\_rendering\_4xslower.avi}

Holograms and three-dimensional renderings of a 0.3 $\times$ 1.1 $\mu$m ellipsoid
breaching an oil-water interface from the aqueous phase. Playback is slower than
real time by a factor of 4.

\bibliography{manuscript-final}

\end{document}